\documentclass[aps,prl,twocolumn,showpacs,preprintnumbers,floatfix,final]{revtex4}

\usepackage{graphicx,amsmath,amssymb,mathrsfs,array,longtable,delarray,verbatim,epstopdf}
\begin{document}

\title{Incipient formation of an electron lattice in a weakly-confined quantum wire}

\author{W. K. Hew}
\author{K. J. Thomas}
\author{M. Pepper}
\author{I. Farrer}
\author{D. Anderson}
\author{G. A. C. Jones}
\author{D. A. Ritchie}

\affiliation{Cavendish Laboratory, J. J. Thomson Avenue, Cambridge CB3 0HE, United Kingdom}


\begin{abstract}
We study the low-temperature transport properties of one-dimensional (1D) quantum wires as the confinement strength $V_{\mathrm{conf}}$ and the carrier density $n_{\mathrm{1D}}$ are varied using a combination of split gates and a top gate in GaAs/AlGaAs heterostructures. At intermediate $V_{\mathrm{conf}}$ and $n_{\mathrm{1D}}$, we observe a jump in conductance to $4e^2/h$, suggesting a double wire. On further reducing $n_{\mathrm{1D}}$, plateau at $2e^2/h$ returns. Our results show beginnings of the formation of an electron lattice in an interacting quasi-1D quantum wire. In the presence of an in-plane magnetic field, mixing of spin-aligned levels of the two wires gives rise to more complex states.
\end{abstract}

 \maketitle

The use of modern semiconductor technology has permitted the fabrication of low-dimensional electron systems, which exhibit quantum transport properties particular to their dimensionality. In the case of a one-dimensional (1D) electron gas,\cite{thornton86, berggren86} the quantisation of ballistic resistance has long been demonstrated in devices with strong confinement,\cite{wharam88, wees88} whereas the r\'egime of weak confinement has largely been overlooked in experimental investigations of quantum wires. However, it has recently been emphasized that interaction can cause a lateral spread of the electron distribution when the confinement weakens.\cite{meyer07} Here, we report on a behaviour of the conductance of quantum wires with varying carrier concentration and confinement, which leads us to suggest a spatial redistribution of the electron system to form a lattice. In previous work, we showed that, at sufficiently low carrier concentrations (but in a strictly 1D r\'egime) the system could enter the spin-incoherent r\'egime.\cite{hew08a} The devices used for the present work operate at higher carrier concentration, where we observe interaction-induced bifurcation of the 1D system into two distinct rows.\footnote{We estimate the Coulomb potential energy due to two neighbouring electrons to be $E_\mathrm{C} \approx 3.5$ meV in the present work, significantly greater than the subband energy spacing of $E_\mathrm{1D} \approx 0.6$ meV, whereas $E_\mathrm{C} \sim E_\mathrm{1D}$ in our previous work.}

By strongly confining electrons to one dimension, the transverse wave-functions are spatially quantized in accordance with what is essentially a  `particle in the box' model. With little or no electron scattering, for example, in short quantum wires, transport is ballistic and the conductance accords well with the predictions of non-interacting theory, where each subband contributes a conductance of $2e^2/h$. The effects of the electron-electron interactions on the quantization are small,\cite{maslov95} except for the structure at $0.7 \times 2e^2/h$,\cite{thomas96} and occasionally that at $0.5 \times 2e^2/h$.\cite{crook06, reilly01, thomas00} A plateau at $0.5 \times 2e^2/h$, attributed to the spin-incoherent r\'egime, has recently been observed.\cite{hew08a} At sufficiently low electron densities $n_{\mathrm{1D}}a_{\mathrm{B}} \ll 1$, where $a_{\mathrm{B}}$ is the effective Bohr radius, Coulomb energy dominates kinetic energy and interactions make possible the formation of a Wigner lattice\cite{wigner38} in which the electrons localise at equidistant sites along a line.\cite{schulz93, klironomos05} For a line of electrons, such ordering can occur simply in the absence of disorder without being accompanied by a change in the topology of the charge distribution. Experimental observations of Wigner crystallization are few and far between, albeit with several notable exceptions.\cite{grimes79, deshpande08}

The theory of electrons in a quantum wire suggests that, when the Coulomb repulsion is sufficiently strong to overcome the confinement potential, the electrons adjust their positions to minimize energy by initially forming a zigzag lattice, a configuration first suggested in the context of electrons on a liquid-helium surface.\cite{chaplik80} In this situation, a variety of phases, including a ferromagnetic state,\cite{matveev04} have been proposed. Both quantum and classical calculations\cite{klironomos07, piacente04} predict that the zigzag will divide as the electron density increases or confinement weakens further, leading to the formation of a lattice, initially of two and then of progressively more rows of electrons, until the system approaches a regular two-dimensional lattice. Because quantisation of conductance is a fundamental property of ballistic transport in 1D, the formation of a spin-polarized system or a transition into rows can be inferred from the conductance characteristics and their evolution with changes in confinement potential and carrier density.

\begin{figure}
\includegraphics[width=1.0\columnwidth]{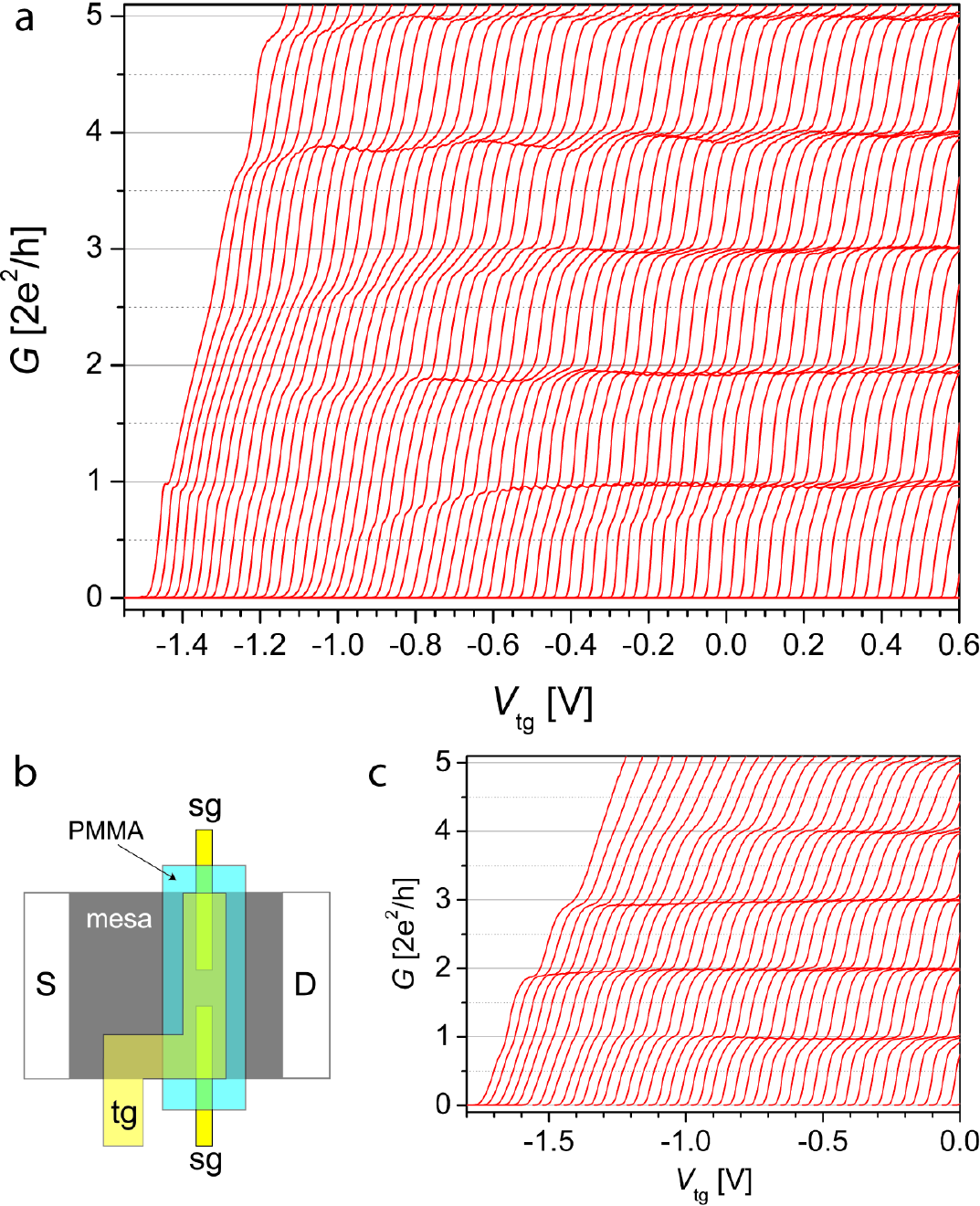}
\caption{a, Conductance traces of sample A measured by sweeping the $V_{\mathrm{tg}}$ at various fixed $V_{\mathrm{sg}}$. Moving right-to-left, $V_{\mathrm{sg}}$ is incremented from $-2$ V to $-0.52$ V in steps of 20 mV, corresponding to a widening of the quantum wire or a weakening of the confining potential. b, A schematic diagram of the device (not to scale) showing the split gates (sg) and top gate (tg) separated by a dielectric layer (PMMA). c, Conductance characteristics of a sample B showing the suppression and disappearance of the plateau at $2e^2/h$ with weakening confinement. All other measurements described in this paper are from sample A.\label{fig1}}
\end{figure}

In order to access these r\'egimes of transport in a quantum wire, independent control over both carrier density and confinement potential is essential, which we achieve by electrostatically depleting a two-dimensional electron gas formed at a GaAs/AlGaAs heterojunction 300 nm deep (mobility $1.85 \times 10^6$ cm$^2$/Vs \& electron density $1.6 \times 10^{11}/$cm$^2$ at 1.5 K) using a pair of split gates and a top gate (Fig.~\,\ref{fig1}b)  separated by an insulating layer of cross-linked PMMA 200 nm thick.\cite{hew08b} The gap between the split gates has a length of 0.4 $\mu$m and a width of 0.7 $\mu$m, and the top gate spans 1 $\mu$m. We measure the two-terminal differential conductance using an excitation voltage of 5 $\mu$V at 33 Hz in a dilution refrigerator with a base temperature of $T \approx 50$ mK. By only weakly biasing the split gates, we can define a wide quantum wire with a shallow confinement potential, controlling the carrier density (of the order of $3 \times 10^5$/cm in the wide limit) with the continuous top-gate. The conductance $G$ can be measured as a function of the carrier density whilst sweeping the top-gate voltage $V_{\mathrm{tg}}$ at a fixed split-gate voltage $V_{\mathrm{sg}}$. The effects described here have been qualitatively reproduced in five other similar samples, as well as in some of a split-gate/mid-gate geometry. 

Figure\,\ref{fig1}a shows a series of conductance traces $G(V_{\mathrm{tg}})$ taken as $V_{\mathrm{sg}}$ is incremented. The confinement potential gets shallower with each successive trace towards the left, which corresponds to a widening of the channel. When the wire is strongly confined (right), the standard ballistic quantisation of conductance is observed with plateaux at multiples of $2e^2/h$. As the wire widens, the lower plateaux weaken and fall below their quantized values. This behaviour is most marked for the first plateau at $2e^2/h$; in the r\'egime of intermediate confinement, this plateau tends to $e^2/h$ and then vanishes, whereupon the conductance jumps directly to the $4e^2/h$ plateau as the wire is populated.  The disappearance of $2e^2/h$ plateau is due to neither thermal averaging nor scattering, for a quantized plateau reappears at $2e^2/h$ when the confinement potential is weakest (far left). Upon further widening, the second and third plateaux (at $4e^2/h$ and $6e^2/h$ respectively) also weaken and become suppressed.

We also note the presence of 0.7 structure in Fig.~\,\ref{fig1}a across a small range of top-gate voltages: at $V_{\mathrm{tg}} \sim 0$ V, the structure occurs at approximately $0.7 \times 2e^2/h$,  but interestingly declines to $0.5 \times 2e^2/h$ at $V_{\mathrm{tg}} \sim -0.5$ V, and then disappears. Application of a parallel magnetic field drives the 0.7 structure to $e^2/h$, confirming its spin-related origin, although it is not clear whether this phenomenon is in fact a manifestation of the ferromagnetic state predicted for a zig-zag chain. However, it is noteworthy that the disappearance of the 0.7 structure coincides precisely with the onset of the descent in the $2e^2/h$ plateau. A jump in conductance straight to $4e^2/h$, without the first plateau, in a manner that is reproducible and unaffected by channel impurities, indicates that the conventional subband model is not appropriate. For normal 1D subbands, the lowest energy state always corresponds to $G=2e^2/h$, and there must be an energy difference between this and the second subband for which $G=4e^2/h$; the absence of an energy difference reflects a breakdown of the non-interacting situation and the direct formation of two rows. Figure\,\ref{fig1}c shows the breakdown of quantization in another sample (sample B) of similar design defined in a different wafer: most of features discussed for sample A are reproduced, albeit more weakly; the robustness of the $4e^2/h$ plateau in the region of breakdown of the first quantized plateau is nonetheless notable.

\begin{figure}
\includegraphics[width=1.0\columnwidth]{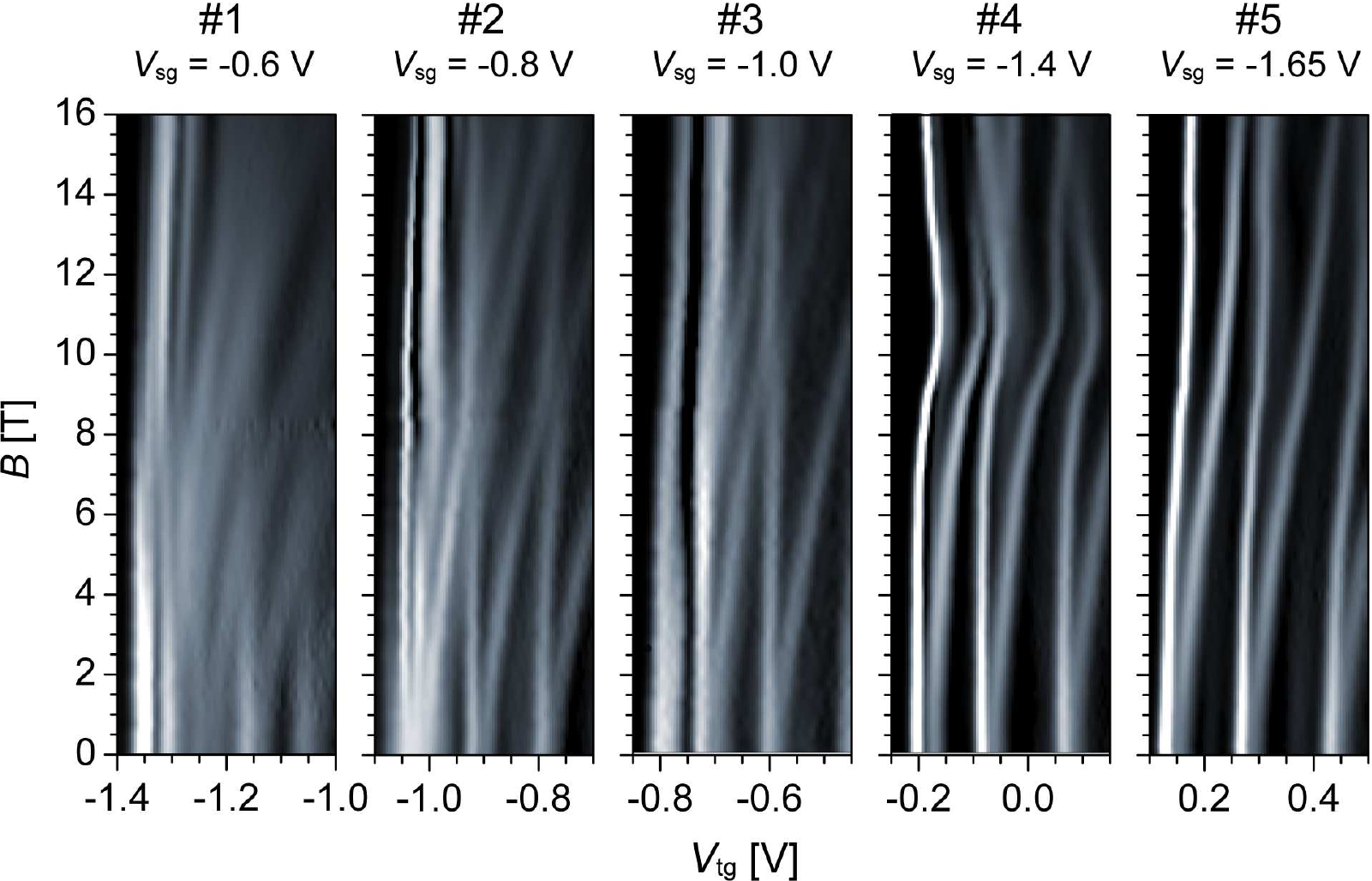}
\caption{Grey-scale plots of transconductance $dG/dV_{\mathrm{tg}}$ as a function of $V_{\mathrm{tg}}$ and $B$ for the last few subbands in sample A. Each panel corresponds to a given $V_{\mathrm{sg}}$, and therefore wire width, which decreases from left to right. Dark and light areas correspond to regions of small and large gradients respectively in $dG/dV_{\mathrm{tg}}$ with respect to $V_{\mathrm{tg}}$. The white lines therefore represent the risers between conductance plateaux, i.e., subband edges, whilst the dark areas represent the plateaux themselves. The re-entrant $2e^2/h$ plateau at far left of Fig.~\,\ref{fig1}a is represented as a thin dark line at $V_{tg} \sim -1.34$ V in the low-field region of Panel \#1. \label{fig2}}
\end{figure}

Figure\,\ref{fig2} shows the evolution of the lowest 1D subbands in a magnetic field at five different confinement strengths. Panels \#4 and \#5 show the expected Zeeman splitting of 1D subbands, represented by the diverging pairs of transconductance peaks in the strong-confinement r\'egime. In panel \#4, we observe a crossing of the 1$\uparrow$ level with 2$\downarrow$ level, and of the 2$\uparrow$ with 3$\downarrow$, when $B$ is approximately 13 T and 15 T respectively. Moving to panel \#3, the corresponding crossings occur at 6 T and 9.5 T, reflecting a continuous reduction in the subband spacing as the confinement weakens. This is obvious from their behaviour at $B = 0$ T, where weakening confinement is accompanied by a reduction in the separation between the first and second spin-degenerate subbands, which culminates in their merging into a four-fold-degenerate `band' in panel \#2, which splits into four branches as $B$ increases.

\begin{figure}
\includegraphics[width=0.8\columnwidth]{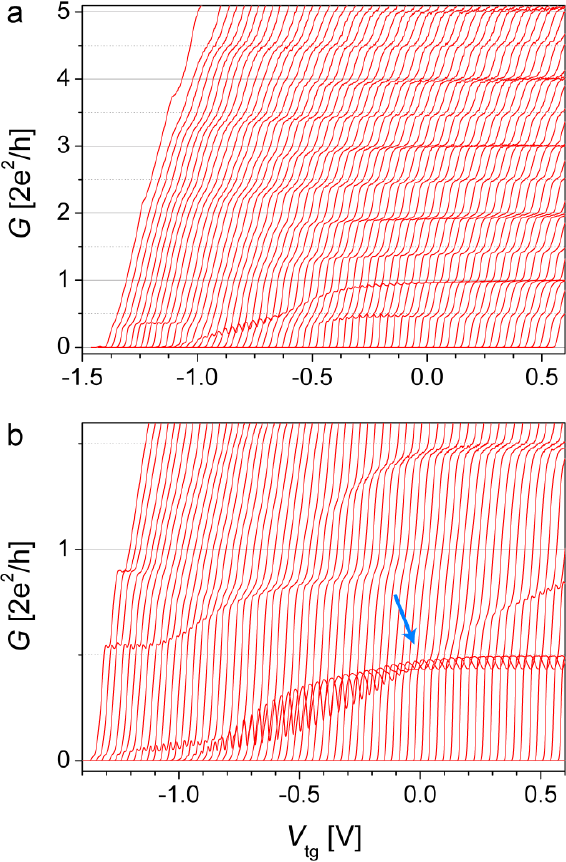}
\caption{a, Conductance traces as in Fig.~\,\ref{fig1}a, but in an in-plane magnetic field $B = 7$ T perpendicular to the transport direction. Well-defined plateaux quantised in multiples of $e^2/h$ are present on the right. The first quantised plateau at $e^2/h$ disappears in the intermediate $V_{\mathrm{tg}}$ r\'egime, coincident with the suppression of $2e^2/h$ plateau. It should be noted that, were it a non-interacting single-mode wire, this would not be possible. The oscillation superimposed on the suppressed plateau can be shown to be due to a mixing of two spin-aligned rows from the corresponding greyscale of this plot. b, A repeat of the measurement shown in a, but with $B = 16$ T. The onset of the oscillations coincides with the incursion of a descending plateau coming from $2e^2/h$, as indicated by the blue arrow. \label{fig3}}
\end{figure}

Figure\,\ref{fig3}a shows the wire characteristics in an in-plane perpendicular magnetic field $B = 7$ T. Zeeman splitting lifts the spin degeneracy, giving rise to additional plateaux at half-integer multiples of $2e^2/h$ in the higher subbands, clearly seen in the strong-confinement r\'egime. However, with weakening confinement, the fall of the $2e^2/h$ plateau now extends to $G = 0$, resulting in the absence of the first plateau ($e^2/h$) across a range of $V_{\mathrm{tg}}$. In Fig.~\,\ref{fig3}b, taken at 16 Tesla, the descending structure becomes distinctly oscillatory below $e^2/h$. The envelope of the oscillations resembles a beat, the rising-falling-rising pattern of the conductance for $G < e^2/h$ attributable to the two lowest (spin-parallel) levels, viz. 1$\downarrow$ and 2$\downarrow$, as they approach one another with weakening confinement. The onset of this pattern coincides with the incursion of a plateau beginning at $2e^2/h$ into the spin-split plateau at $e^2/h$---both plateaux correspond to spin-down ($\downarrow$) levels. The same effect is also observed, albeit more weakly, at $B = 7$ T in Fig.~\,\ref{fig3}a, where a small oscillation can be seen on the descending plateau originating at $2e^2/h$ as soon as it crosses below the plateau at $e^2/h$. We speculate that this effect arises at the threshold of the electron system bifurcating, when the single- and double-row formations are very nearly equal in energy, and switching between these two configurations produces the beat-like envelope. In zero field, the subbands are spin degenerate and the structure therefore absent.

\begin{figure}
\includegraphics[width=1.0\columnwidth]{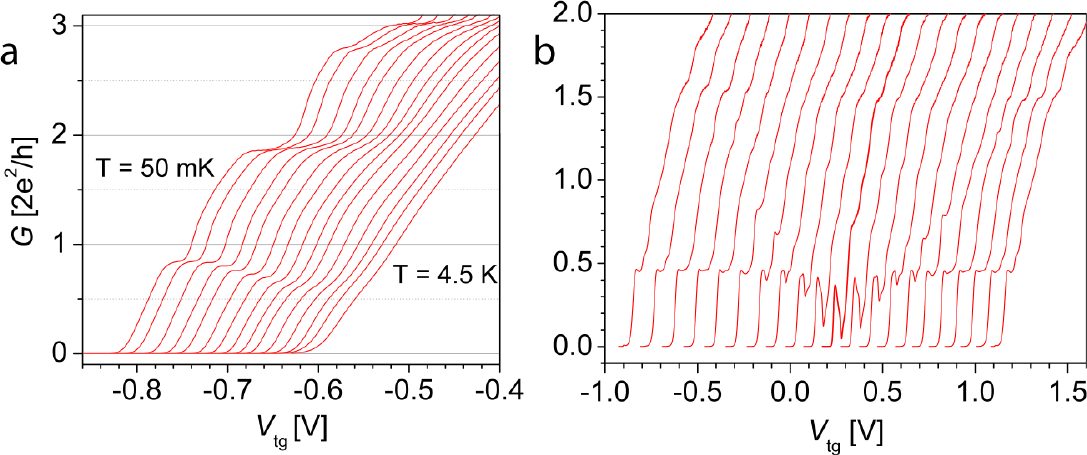}
\caption{a, Conductance traces taken at $V_{\mathrm{sg}} = -1$ V showing the evolution of the declining $2e^2/h$ plateau with increasing temperature. From left to right, $T_{\mathrm{lattice}}$ = 0.05, 0.1, 0.2, 0.3, 0.4, 0.5, 0.6, 0.8, 1.0, 1.2, 1.45, 1.95, 2.65, 3.5, 4.5 K. [Traces have been offset from the 50~mK trace for clarity.] b, A single trace from Fig.~\,\ref{fig3}b showing the behaviour of the oscillatory structure as the channel is shifted laterally by differentially biasing the split gates. (Successive traces are offset, the bold trace corresponding to equal bias on the split gates.) \label{fig4}}
\end{figure}

Figure\,\ref{fig4}a shows the temperature dependence of a typical trace with declining $2e^2/h$ plateau seen in Fig. 1c. Its conductance is suppressed towards $e^2/h$ as the temperature is increased, until it finally disappears at $T \sim 3.5$ K due to thermal averaging. Note that, although the higher plateaux at $4e^2/h$ and $6e^2/h$ similarly weaken with increasing temperature, their conductance values remain constant. It is likely that the increase in temperature drives the system into the spin-incoherent r\'egime, \cite{fiete07} which depends on thermal energy dominating exchange energy. Nevertheless, the unusual temperature dependence rules out Coulomb blockade and other disorder effects as the cause of the suppression of the $2e^2/h$ plateau.

To further test the origin of the oscillatory structure in a magnetic field, the channel was laterally shifted by differentially biasing the split gates. As seen in Fig.~\,\ref{fig4}b, the structure is initially stable and then disappears symmetrically in either direction of shift, the conductance levelling off at $e^2/h$, which suggests the destruction of row formation and the return of a single subband as expected for strong confinement. The symmetry of the traces with respect to bias polarity shows the channel is free of impurities and other significant imperfections. That disorder is largely absent therefore explains why ballistic quantisation of conductance can be observed in the double-row configuration without evidence of pinning.

The decline of the $2e^2/h$ plateau towards $e^2/h$ with increasing temperature or decreasing density reflects a transition into an strongly-interacting quasi-1D transport r\'egime. Whilst zigzag configuration has been theoretically predicted\cite{klironomos07} in this transition, we cannot experimentally verify the existence of such a spatial distribution of electrons, likely though it may be as a precursor to the formation of rows. Moving to the left (in Fig.~\,\ref{fig1}a), as the channel widens and the density decreases, the chain further staggers, whereupon the conductance jumps to $4e^2/h$, reflecting the formation of two spin-degenerate rows, each exhibiting ballistic 1D transport. In two dimensions, a Wigner crystal should be pinned by any disorder,\cite{chitra01} on the basis of which it appears likely that the double-row formation reported here does not represent a rigid crystal. The reappearance of $2e^2/h$ plateau at the lowest densities (far left) most likely represents the return of the normal single-wire characteristics as the system passes beyond the conditions favouring the double-row configuration.\cite{meyer07}

Previous experimental efforts in one-dimensional transport have largely focussed on a 1D electron gas produced by strong confinement. The flexibility of our devices has made it possible to explore the opposite limit, that of weak confinement, in a systematic fashion. By studying the conductance plateaux in this limit, we have shown the breakdown of a quasi-1D system of electrons defined by spatial quantisation. The weakening of this quantisation indicates that the interaction has driven the system into a new configuration of degenerate rows, which may be associated with jumps of conductance in multiples of $4e^2/h$. We have only discussed the principal features of this new r\'egime of weakly-confined electron gases, of which double-row formation and the onset and disappearance of spin polarisation are but a part. Further work is required to fully understand the mechanism of multiple row formation and other exotic phases predicted in this r\'egime. However, the formation of a double row and the breakdown of the subband model of structure is a necessary first step towards a more complex lattice.

This work was supported by the Engineering and Physical Sciences Research Council. WKH acknowledges the Cambridge Commonwealth Trust and KJT the Royal Society. We thank J. Meyer and N. Cooper for useful discussions.

\bibliographystyle{apsrev}

\begin{thebibliography}{23}
\expandafter\ifx\csname natexlab\endcsname\relax\def\natexlab#1{#1}\fi
\expandafter\ifx\csname bibnamefont\endcsname\relax
  \def\bibnamefont#1{#1}\fi
\expandafter\ifx\csname bibfnamefont\endcsname\relax
  \def\bibfnamefont#1{#1}\fi
\expandafter\ifx\csname citenamefont\endcsname\relax
  \def\citenamefont#1{#1}\fi
\expandafter\ifx\csname url\endcsname\relax
  \def\url#1{\texttt{#1}}\fi
\expandafter\ifx\csname urlprefix\endcsname\relax\def\urlprefix{URL }\fi
\providecommand{\bibinfo}[2]{#2}
\providecommand{\eprint}[2][]{\url{#2}}

\bibitem[{\citenamefont{Thornton et~al.}(1986)}]{thornton86}
\bibinfo{author}{\bibfnamefont{T.~J.} \bibnamefont{Thornton}},
  \bibinfo{author}{\bibfnamefont{M.}~\bibnamefont{Pepper}},
  \bibinfo{author}{\bibfnamefont{H.}~\bibnamefont{Ahmed}},
  \bibinfo{author}{\bibfnamefont{D.}~\bibnamefont{Andrews}}, \bibnamefont{and}
  \bibinfo{author}{\bibfnamefont{G.~J.} \bibnamefont{Davies}},
  \bibinfo{journal}{Phys.\ Rev.\ Lett.} \textbf{\bibinfo{volume}{56}},
  \bibinfo{pages}{1198} (\bibinfo{year}{1986}).

\bibitem[{\citenamefont{Berggren et~al.}(1986)}]{berggren86}
\bibinfo{author}{\bibfnamefont{K.~F.} \bibnamefont{Berggren}},
  \bibinfo{author}{\bibfnamefont{T.~J.} \bibnamefont{Thornton}},
  \bibinfo{author}{\bibfnamefont{D.~J.} \bibnamefont{Newson}},
  \bibnamefont{and} \bibinfo{author}{\bibfnamefont{M.}~\bibnamefont{Pepper}},
  \bibinfo{journal}{Phys.\ Rev.\ Lett.} \textbf{\bibinfo{volume}{57}},
  \bibinfo{pages}{1769} (\bibinfo{year}{1986}).

\bibitem[{\citenamefont{Wharam et~al.}(1988)}]{wharam88}
\bibinfo{author}{\bibfnamefont{D.~A.} \bibnamefont{Wharam}},
\bibinfo{author}{\bibfnamefont{T.~J.} \bibnamefont{Thornton}},
\bibinfo{author}{\bibfnamefont{R.} \bibnamefont{Newbury}},
  \bibinfo{author}{\bibfnamefont{M.}~\bibnamefont{Pepper}},
  \bibinfo{author}{\bibfnamefont{H.}~\bibnamefont{Ahmed}},
  \bibinfo{author}{\bibfnamefont{J.~E.~F.}~\bibnamefont{Frost}},
  \bibinfo{author}{\bibfnamefont{D.~G.}~\bibnamefont{Hasko}},
  \bibinfo{author}{\bibfnamefont{D.~C.}~\bibnamefont{Peacock}},
  \bibinfo{author}{\bibfnamefont{D.~A.} \bibnamefont{Ritchie}}, \bibnamefont{and}
  \bibinfo{author}{\bibfnamefont{G.~A.~C.} \bibnamefont{Jones}},
  \bibinfo{journal}{J.\ Phys.\ C: Solid State Phys.} \textbf{\bibinfo{volume}{21}},
  \bibinfo{pages}{L209} (\bibinfo{year}{1988}).

\bibitem[{\citenamefont{van Wees et~al.}(2007)}]{wees88}
\bibinfo{author}{\bibfnamefont{B.~J.} \bibnamefont{van Wees}},
  \bibinfo{author}{\bibfnamefont{H.}~\bibnamefont{van Houten}},
  \bibinfo{author}{\bibfnamefont{C.~W.~J.}~\bibnamefont{Beenakker}},
  \bibinfo{author}{\bibfnamefont{J.~G.}~\bibnamefont{Williamson}},
  \bibinfo{author}{\bibfnamefont{L.~P.}~\bibnamefont{Kouwenhoven}},
  \bibinfo{author}{\bibfnamefont{D.}~\bibnamefont{van der Marel}}, \bibnamefont{and}
  \bibinfo{author}{\bibfnamefont{C.~T.} \bibnamefont{Foxon}},
  \bibinfo{journal}{Phys.\ Rev.\ Lett.} \textbf{\bibinfo{volume}{60}},
  \bibinfo{pages}{848} (\bibinfo{year}{1988}).

\bibitem[{\citenamefont{Meyer et~al.}(2007)}]{meyer07}
\bibinfo{author}{\bibfnamefont{J.~S.} \bibnamefont{Meyer}},
\bibinfo{author}{\bibfnamefont{K.~A.} \bibnamefont{Matveev}}, \bibnamefont{and}
\bibinfo{author}{\bibfnamefont{A.~I.} \bibnamefont{Larkin}},
  \bibinfo{journal}{Phys.\ Rev.\ Lett.} \textbf{\bibinfo{volume}{98}},
  \bibinfo{pages}{126404} (\bibinfo{year}{2007}).

\bibitem[{\citenamefont{Hew et~al.}(1996)}]{hew08a}
\bibinfo{author}{\bibfnamefont{W.~K.} \bibnamefont{Hew}},
\bibinfo{author}{\bibfnamefont{K.~J.} \bibnamefont{Thomas}},
  \bibinfo{author}{\bibfnamefont{M.}~\bibnamefont{Pepper}},
  \bibinfo{author}{\bibfnamefont{I.}~\bibnamefont{Farrer}},
  \bibinfo{author}{\bibfnamefont{D.}~\bibnamefont{Anderson}},
  \bibinfo{author}{\bibfnamefont{G.~A.~C.} \bibnamefont{Jones}}, \bibnamefont{and}
  \bibinfo{author}{\bibfnamefont{D.~A.} \bibnamefont{Ritchie}},
  \bibinfo{journal}{Phys.\ Rev.\ Lett.} \textbf{\bibinfo{volume}{101}},
  \bibinfo{pages}{036801} (\bibinfo{year}{2008}).

\bibitem[{\citenamefont{Maslov and Stone}(1995)}]{maslov95}
\bibinfo{author}{\bibfnamefont{D.~L.} \bibnamefont{Maslov}} \bibnamefont{and}
  \bibinfo{author}{\bibfnamefont{M.}~\bibnamefont{Stone}},
  \bibinfo{journal}{Phys.\ Rev.\ B} \textbf{\bibinfo{volume}{52}},
  \bibinfo{pages}{R5539} (\bibinfo{year}{1995}).

\bibitem[{\citenamefont{Thomas et~al.}(1996)}]{thomas96}
\bibinfo{author}{\bibfnamefont{K.~J.} \bibnamefont{Thomas}},
  \bibinfo{author}{\bibfnamefont{J.~T.} \bibnamefont{Nicholls}},
  \bibinfo{author}{\bibfnamefont{M.~Y.} \bibnamefont{Simmons}},
  \bibinfo{author}{\bibfnamefont{M.}~\bibnamefont{Pepper}},
  \bibinfo{author}{\bibfnamefont{D.~R.} \bibnamefont{Mace}}, \bibnamefont{and}
  \bibinfo{author}{\bibfnamefont{D.~A.} \bibnamefont{Ritchie}},
  \bibinfo{journal}{Phys.\ Rev.\ Lett.} \textbf{\bibinfo{volume}{77}},
  \bibinfo{pages}{135} (\bibinfo{year}{1996}).

\bibitem[{\citenamefont{Crook et~al.}(2006)\citenamefont{Crook, Prance, Thomas,
  Chorley, Farrer, Ritchie, Pepper, and Smith}}]{crook06}
\bibinfo{author}{\bibfnamefont{R.}~\bibnamefont{Crook}},
  \bibinfo{author}{\bibfnamefont{J.}~\bibnamefont{Prance}},
  \bibinfo{author}{\bibfnamefont{K.~J.} \bibnamefont{Thomas}},
  \bibinfo{author}{\bibfnamefont{S.~J.} \bibnamefont{Chorley}},
  \bibinfo{author}{\bibfnamefont{I.}~\bibnamefont{Farrer}},
  \bibinfo{author}{\bibfnamefont{D.~A.} \bibnamefont{Ritchie}},
  \bibinfo{author}{\bibfnamefont{M.}~\bibnamefont{Pepper}}, \bibnamefont{and}
  \bibinfo{author}{\bibfnamefont{C.~G.} \bibnamefont{Smith}},
  \bibinfo{journal}{Science} \textbf{\bibinfo{volume}{312}},
  \bibinfo{pages}{1359} (\bibinfo{year}{2006}).

\bibitem[{\citenamefont{Reilly et~al.}(1995)}]{reilly01}
  \bibinfo{author}{\bibfnamefont{D.~J.}~\bibnamefont{Reilly}},
  \bibinfo{author}{\bibfnamefont{G.~R.} \bibnamefont{Facer}},
  \bibinfo{author}{\bibfnamefont{A.~S.} \bibnamefont{Dzurak}},
  \bibinfo{author}{\bibfnamefont{B.~E.}~\bibnamefont{Kane}},
  \bibinfo{author}{\bibfnamefont{R.~G.} \bibnamefont{Clark}},
  \bibinfo{author}{\bibfnamefont{P.~J.} \bibnamefont{Stiles}},
  \bibinfo{author}{\bibfnamefont{J.~L.} \bibnamefont{O'Brien}},
  \bibinfo{author}{\bibfnamefont{N.~E.}~\bibnamefont{Lumpkin}},
  \bibinfo{author}{\bibfnamefont{L.~N.} \bibnamefont{Pfeiffer}}, \bibnamefont{and}
  \bibinfo{author}{\bibfnamefont{K.~W.} \bibnamefont{West}},
  \bibinfo{journal}{Phys.\ Rev.\ B} \textbf{\bibinfo{volume}{63}},
  \bibinfo{pages}{121311(R)} (\bibinfo{year}{2001}).

\bibitem[{\citenamefont{Thomas et~al.}(2000)}]{thomas00}
\bibinfo{author}{\bibfnamefont{K.~J.} \bibnamefont{Thomas}},
  \bibinfo{author}{\bibfnamefont{J.~T.} \bibnamefont{Nicholls}},
  \bibinfo{author}{\bibfnamefont{M.}~\bibnamefont{Pepper}},
  \bibinfo{author}{\bibfnamefont{W.~R.} \bibnamefont{Tribe}},
  \bibinfo{author}{\bibfnamefont{M.~Y.} \bibnamefont{Simmons}}, \bibnamefont{and}
  \bibinfo{author}{\bibfnamefont{D.~A.} \bibnamefont{Ritchie}},
  \bibinfo{journal}{Phys.\ Rev.\ B} \textbf{\bibinfo{volume}{61}},
  \bibinfo{pages}{R13365} (\bibinfo{year}{2000}).

\bibitem[{\citenamefont{Wigner}(2007)}]{wigner38}
\bibinfo{author}{\bibfnamefont{E.~P.} \bibnamefont{Wigner}},
  \bibinfo{journal}{Trans.\ Faraday Soc.} \textbf{\bibinfo{volume}{34}},
  \bibinfo{pages}{678} (\bibinfo{year}{1938}).

\bibitem[{\citenamefont{Schulz}(1993)}]{schulz93}
\bibinfo{author}{\bibfnamefont{H.~J.} \bibnamefont{Schulz}},
  \bibinfo{journal}{Phys.\ Rev.\ Lett.} \textbf{\bibinfo{volume}{71}},
  \bibinfo{pages}{1864} (\bibinfo{year}{1993}).

\bibitem[{\citenamefont{Klironomos et~al.}(2005)}]{klironomos05}
\bibinfo{author}{\bibfnamefont{A.~D.} \bibnamefont{Klironomos}},
\bibinfo{author}{\bibfnamefont{R.~R.} \bibnamefont{Ramazashvili}}, \bibnamefont{and}
\bibinfo{author}{\bibfnamefont{K.~A.} \bibnamefont{Matveev}},
  \bibinfo{journal}{Phys.\ Rev.\ B} \textbf{\bibinfo{volume}{72}},
  \bibinfo{pages}{195343} (\bibinfo{year}{2005}).

\bibitem[{\citenamefont{Grimes et~al.}(2004)}]{grimes79}
\bibinfo{author}{\bibfnamefont{C.~C.} \bibnamefont{Grimes}} \bibnamefont{and}
\bibinfo{author}{\bibfnamefont{G.} \bibnamefont{Adams}},
  \bibinfo{journal}{Phys.\ Rev.\ Lett.}
  \textbf{\bibinfo{volume}{42}}, \bibinfo{pages}{795}
  (\bibinfo{year}{1979}).

\bibitem[{\citenamefont{Deshpande et~al.}(2008)}]{deshpande08}
\bibinfo{author}{\bibfnamefont{V.~V.} \bibnamefont{Deshpande}} \bibnamefont{and}
\bibinfo{author}{\bibfnamefont{M.} \bibnamefont{Bockrath}},
  \bibinfo{journal}{Nature Physics}
  \textbf{\bibinfo{volume}{4}}, \bibinfo{pages}{314}
  (\bibinfo{year}{2008}).

\bibitem[{\citenamefont{Chaplik}(1980)}]{chaplik80}
\bibinfo{author}{\bibfnamefont{A.~V.} \bibnamefont{Chaplik}},
  \bibinfo{journal}{JETP Lett.} \textbf{\bibinfo{volume}{31}},
  \bibinfo{pages}{252} (\bibinfo{year}{1980}).

\bibitem[{\citenamefont{Matveev}(2004)}]{matveev04}
\bibinfo{author}{\bibfnamefont{K.~A.} \bibnamefont{Matveev}},
  \bibinfo{journal}{Phys.\ Rev.\ Lett.} \textbf{\bibinfo{volume}{92}},
  \bibinfo{pages}{106801} (\bibinfo{year}{2004}).

\bibitem[{\citenamefont{Klironomos et~al.}(2007)}]{klironomos07}
\bibinfo{author}{\bibfnamefont{A.~D.} \bibnamefont{Klironomos}},
\bibinfo{author}{\bibfnamefont{J.~S.} \bibnamefont{Meyer}},
\bibinfo{author}{\bibfnamefont{T.} \bibnamefont{Hikihara}}, \bibnamefont{and}
\bibinfo{author}{\bibfnamefont{K.~A.} \bibnamefont{Matveev}},
  \bibinfo{journal}{Phys.\ Rev.\ B} \textbf{\bibinfo{volume}{76}},
  \bibinfo{pages}{075302} (\bibinfo{year}{2007}).
  
\bibitem[{\citenamefont{Piacente et~al.}(2004)}]{piacente04}
\bibinfo{author}{\bibfnamefont{G.} \bibnamefont{Piacente}},
\bibinfo{author}{\bibfnamefont{I.~V.} \bibnamefont{Schweigert}},
\bibinfo{author}{\bibfnamefont{J.~J.} \bibnamefont{Betouras}}, \bibnamefont{and}
\bibinfo{author}{\bibfnamefont{F.~M.} \bibnamefont{Peeters}},
  \bibinfo{journal}{Phys.\ Rev.\ B} \textbf{\bibinfo{volume}{69}},
  \bibinfo{pages}{045324} (\bibinfo{year}{2004}).

\bibitem[{\citenamefont{Hew et~al.}(1996)}]{hew08b}
\bibinfo{author}{\bibfnamefont{W.~K.} \bibnamefont{Hew}},
\bibinfo{author}{\bibfnamefont{K.~J.} \bibnamefont{Thomas}},
  \bibinfo{author}{\bibfnamefont{I.}~\bibnamefont{Farrer}},
  \bibinfo{author}{\bibfnamefont{D.}~\bibnamefont{Anderson}},
  \bibinfo{author}{\bibfnamefont{D.~A.} \bibnamefont{Ritchie}}, \bibnamefont{and}
  \bibinfo{author}{\bibfnamefont{M.}~\bibnamefont{Pepper}},
  \bibinfo{journal}{Physica E} \textbf{\bibinfo{volume}{40}},
  \bibinfo{pages}{1645} (\bibinfo{year}{2008}).

\bibitem[{\citenamefont{Fiete}(2007)}]{fiete07}
\bibinfo{author}{\bibfnamefont{G.~A.} \bibnamefont{Fiete}},
  \bibinfo{journal}{Rev. Mod. Phys.} \textbf{\bibinfo{volume}{79}},
  \bibinfo{pages}{801} (\bibinfo{year}{2007}).

\bibitem[{\citenamefont{Deshpande et~al.}(2008)}]{chitra01}
\bibinfo{author}{\bibfnamefont{R.} \bibnamefont{Chitra}},
\bibinfo{author}{\bibfnamefont{T.} \bibnamefont{Giamarchi}} \bibnamefont{and}
\bibinfo{author}{\bibfnamefont{P.} \bibnamefont{Le~Doussal}},
  \bibinfo{journal}{Phys.\ Rev.\ B} \textbf{\bibinfo{volume}{65}},
  \bibinfo{pages}{035312} (\bibinfo{year}{2001}).

\end{thebibliography}

\end{document}